\documentclass[journal=jpcbfk,manuscript=article]{achemso}

\usepackage[version=3]{mhchem} % Formula subscripts using \ce{}
\usepackage{siunitx}
\usepackage[T1]{fontenc}

\usepackage{pdfpages}

\author{Tim Schrader}
\affiliation[FSU Jena]
{Otto Schott Institute of Materials Research (OSIM), Faculty of Physics and Astronomy, Friedrich Schiller University Jena, L\"obdergraben 32, 07743 Jena, Germany}
\author{Alen Shaji}
\affiliation[FSU Jena]
{Otto Schott Institute of Materials Research (OSIM), Faculty of Physics and Astronomy, Friedrich Schiller University Jena, L\"obdergraben 32, 07743 Jena, Germany}
\author{Christin David}
\affiliation{University of Applied Sciences Landshut, Am Lurzenhof 1, 84036 Landshut}
\author{Eva Perlt}
\affiliation[FSU Jena]
{Otto Schott Institute of Materials Research (OSIM), Faculty of Physics and Astronomy, Friedrich Schiller University Jena, L\"obdergraben 32, 07743 Jena, Germany}
\email{eva.von.domaros@uni-jena.de}

\title{Theoretical investigation on the formation of ethylene on a gold surface.}

\begin{document}

\begin{abstract}
The targeted and efficient \ce{CO2} reduction remains an appealing option to capture \ce{CO2} from the atmosphere and transform it into value-added chemicals.
The formation of methylene and subsequent dimerization to ethylene is one possible step in the rather complex pathway.
These reactions typically occur on (catalytic) surfaces. It is therefore critical to gain a  deeper understanding of the role of the surface.
We study the reaction pathway of the dimerization on gold surfaces with varying layer thickness.
The presence of support layers has a significant influence on the reaction coordinate, the energetic profile and finally on the adsorption energy of the product on the surface with interesting consequences for the role of nanostructured gold surfaces for \ce{CO2} upconversion.
\end{abstract}
%%%%%%%%%%%%%%%%%%%%%%%%%%%%%%%%%%%%%%%%%%%%%%%%%%%%%%%%%%

%%%%%%%		 Main Text			%%%%%%% 

\section*{Introduction}
\label{introduction}
Within global carbon dioxide (\ce{CO2}) emissions still growing each year, it is increasingly abundant in the atmosphere and as a greenhouse gas the major component relevant for global warming.\cite{friedlingstein_global_2025}
Thus, various strategies of carbon capture and storage are being researched and pursued in an effort to eliminate \ce{CO2} from the atmosphere.\cite{gowd_economic_2023}
An even more desirable approach is the conversion of \ce{CO2} to value-added chemicals by reduction and upconversion.\cite{da_silva_freitas_electrocatalytic_2021, li_heterogeneous_2015}
This, however, requires suitable catalytic materials and energy conversion strategies.

The class of carbon dioxide reduction reactions %(\ce{CO2}RR)
includes complex pathways, in which \ce{CO2} is reduced to form products including methanol, methane, and others.\cite{ao_theoretical_2020,li_simply_2020}
If these \ce{C1} species couple, they form \ce{C2} species or even longer carbon chains.\cite{wang_insight_2022,fusco_advances_2024}
Therefore, this coupling step is crucial in carbon upconversion to achieve high-value carbon products.
Currently, these reactions are mainly performed on copper.\cite{peterson_how_2010,kuhl_new_2012,wei_improving_2022,fusco_advances_2024}
However, other materials, especially when used as nanoparticles, may offer alternatives.
Fusco et al. studied plasmonic photocatalysis on small gold clusters with special focus on the presence of adsorbed hydroxide ions.\cite{fusco_ab_2024}

Gold is a particularly interesting material for catalytic surfaces, since gold na\-no\-par\-ti\-cles can enhance the \ce{CO2} reduction process via local plasmonic field effects.\cite{fusco_ab_2024}
These nanoparticles can form large and complex dendritic structures\cite{fusco_cathodoluminescence_2023} which makes it a challenge to fully describe them computationally. Nanoscale edges of branches in the micro\-meter-sized structures play an important role in the formation of hot-spots leading to local plasmon-assisted enhancement of the target reaction.
\cite{fusco_advances_2024}

In this work, we focus on the C-C bond formation in the reaction of two methylene units to form ethylene.
\begin{equation} \label {eq1}
			\ce{2 CH2 -> H2C=CH2}
\end{equation}
The treatment of an entire dendritic structure is computationally infeasible nor considered necessary as local effects are the main drivers in plasmonic photocatalysis.
Therefore, in this study we focus on the effect of the gold layer thickness, comparing the reaction on a monolayer and on three layers of gold to each other.
We apply the state-of-the-art continuous-fast multipole method, which allows applying a local basis rather than a plane wave basis---which is computationally more feasible to cover these molecular reactions---in a 2D periodic setting.

To elucidate the reaction mechanism and energetics of the adsorption and transformation of \ce{CH2} on gold surfaces, a systematic computational study was conducted. 

\section*{Computational Details}
\label{computational_details}
To optimize adsorbed reactants and prevent spontaneous combination during the geometry optimization, a single \ce{CH2} molecule adsorbed on a 2D periodic gold (111) surface was optimized. The surface lattice was redefined with
\begin{equation}
\begin{pmatrix}
\vec{u}\\
\vec{v}
\end{pmatrix} =
\begin{bmatrix}
-3 & -1 \\
0 & 3
\end{bmatrix}
\begin{pmatrix}
\vec{u}_{111}\\
\vec{v}_{111}
\end{pmatrix},
\end{equation}
resulting in the following cell parameters: $\left|\vec{u}\right|=7.629801\ \si{ \angstrom}$, $\left|\vec{v}\right|=8.651381\ \si{ \angstrom}$ and $\theta=79.10661\si{\degree}$.
This maximizes the distance between adsorbed molecules while keeping the cell size small.
The gold positions were constrained during the optimization process to simulate a rigid surface environment.
This cell was doubled in one dimension to accommodate two \ce{CH2} mole\-cules on a fixed 6 by 3 gold cell, representing the reactant state. For the product state, one molecule of ethylene (\ce{H_2C=CH_2}) was placed on the fixed 6 by 3 gold cell and its structure optimized.
In order to estimate the effect of the number of gold layers, the calculations were performed on a single layer of gold as well as on three layers.

All calculations were performed using the riper module of the Turbomole 7.9 program package.\cite{lazarski_density_2015,lazarski_density_2016,franzke_turbomole_2023}
The DFT calculations were performed using the Perdew-Burke-Ernzerhof (PBE) functional \cite{perdew_accurate_1992,perdew_generalized_1996} with a Grimme dispersion correction (DFT-D3)\cite{grimme_consistent_2010} with Becke-Johnson damping.\cite{grimme_effect_2011}
The basis set chosen for these calculations was pob-TZVP-rev2 \cite{vilela_oliveira_bssecorrection_2019} for carbon and hydrogen.
Except for the valence electrons, effective core potentials (ECP78SDF) \cite{fuentealba_reliability_1983} were applied to model electrons of gold atoms together with the basis set ECP78SDF\_GUESS to facilitate convergence of the electronic structure calculation and reduce computational resources.

To model the reaction path of the formation of ethylene, constrained optimizations along the reaction path using quadratic potentials as implemented in Turbomole's module \textsc{woelfling} were applied.\cite{woelfling}
Along the reaction path, 14 structures were generated for both systems with one and three supporting gold layers, respectively.
Therein, the gold atoms' positions were kept in place.

The adiabatic interaction energies of the adsorbed species were calculated with respect to isolated fragments in their relaxed geometries.
The adsorption energies of the product and reactants were calculated with the Counterpoise correction method to account for the basis set superposition error.\cite{boys_calculation_1970}

\section*{Results and Discussion}
\label{results_discussion}
\begin{figure}[b]
\begin{center}
\includegraphics[width=8.6cm]{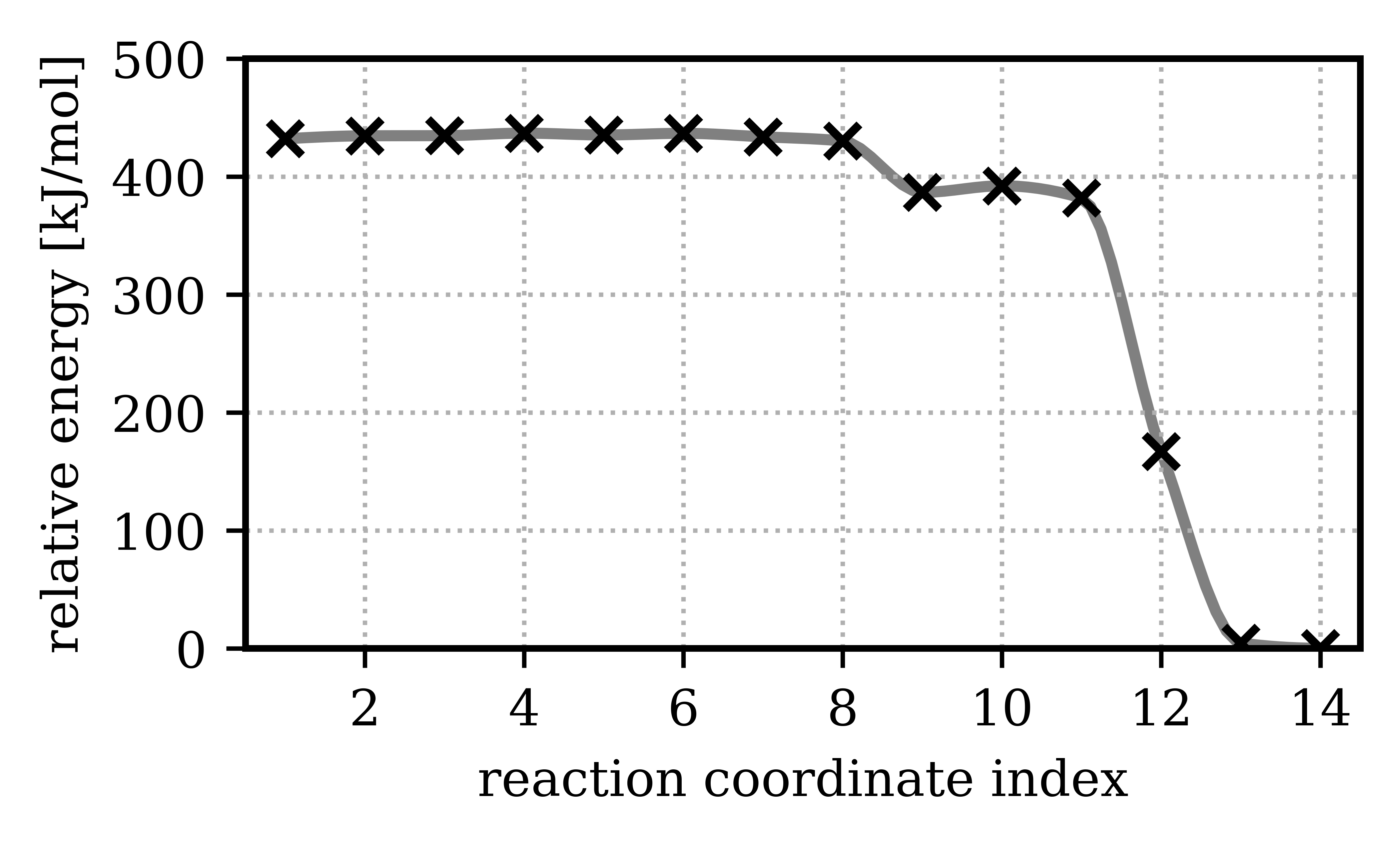}
\caption{Relative energy along the reaction path on one layer of gold.}
\label{energyOverReactionOn1Layers}
\end{center}
\end{figure}

\begin{figure}[b]
\begin{center}
\includegraphics[width=8.6cm]{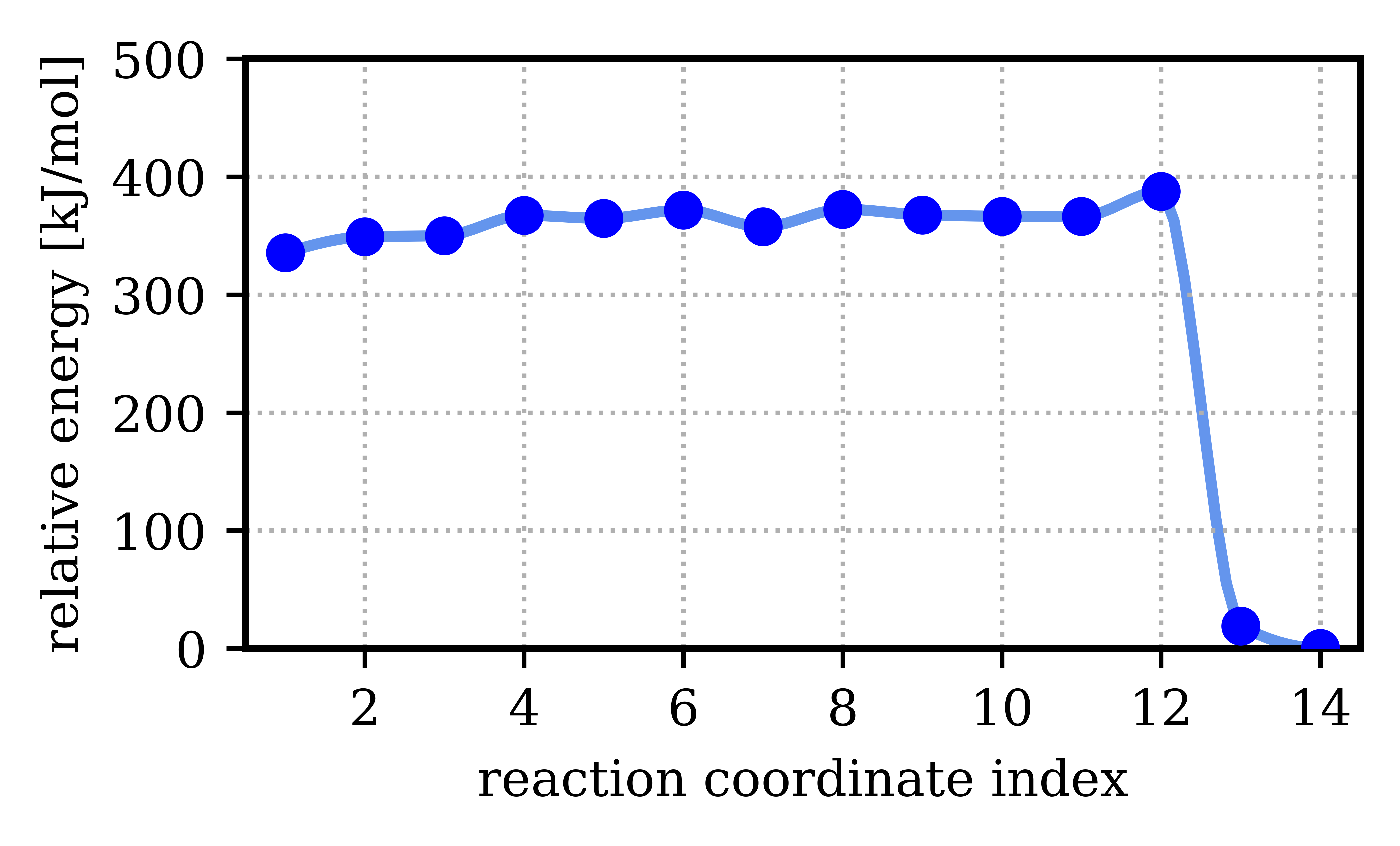}
\caption{Relative energy along the reaction path on three layers of gold.}
\label{energyOverReactionOn3Layers}
\end{center}
\end{figure}

The energy profile along the reaction is studied, which is shown in Figures~\ref{energyOverReactionOn1Layers} and \ref{energyOverReactionOn3Layers} for one and three supporting gold layers, respectively.
Adsorption energies $E_{\mathrm{ads}}$, reaction energies $\Delta E_{\mathrm{react}}$ as well as activation energies $E_a$ for each system are listed in Table~\ref{tab:energies}.
The stable product state is chosen as the energy reference.
At first glance, both curves show a similar behavior with an almost constant energy over the first part of the reaction followed by a steep decrease in energy around the reaction coordinate with index 12.
This is in agreement with the reaction happening in two steps: the diffusion of both educt fragments towards each other on the surface followed by the bond formation and relaxation of the product in the second part, which goes along with a steep decrease of the energy in both scenarios, see Figures~\ref{energyOverReactionOn1Layers} and \ref{energyOverReactionOn3Layers}.
However, both scenarios show subtle differences.
Firstly, the overall reaction energy has a larger absolute value on the monolayer (\SI{-432.06}{\kilo\joule\per\mole}) as compared to the thicker system (\SI{-335.50}{\kilo\joule\per\mole}).
Furthermore, both reactant and product species are more strongly bound to the monolayer surface than to the three-layered system.
In both cases, the products are very loosely bound, which can be attributed to the formation of a $\pi$ bond and the resulting repulsion to the surface electrons.
Consequently, products likely desorb easily.

\begin{table}[t]
    \caption{Characteristic energies of the reaction \ce{2CH2 -> C2H4} on one and three layers of gold, respectively.}
    \label{tab:energies}
    \begin{tabular}{lrr}
    \hline\hline
        energy [\si{\kilo\joule\per\mole}]  & 1 layer & 3 layers \\
        \hline
        $E_{\mathrm{ads}}$ (reactants) & $-137.70$ & $-115.49$ \\
        $E_{\mathrm{ads}}$ (products)  &  $-46.00$ &   $-5.04$ \\
        $\Delta E_{\mathrm{react}}$    & $-432.06$ & $-335.50$\\
        $E_{a}$                        &    $4.78$ &   $52.10$ \\
        \hline\hline
    \end{tabular}
\end{table}

\begin{figure*}
    \centering
    \includegraphics[width=0.49\linewidth]{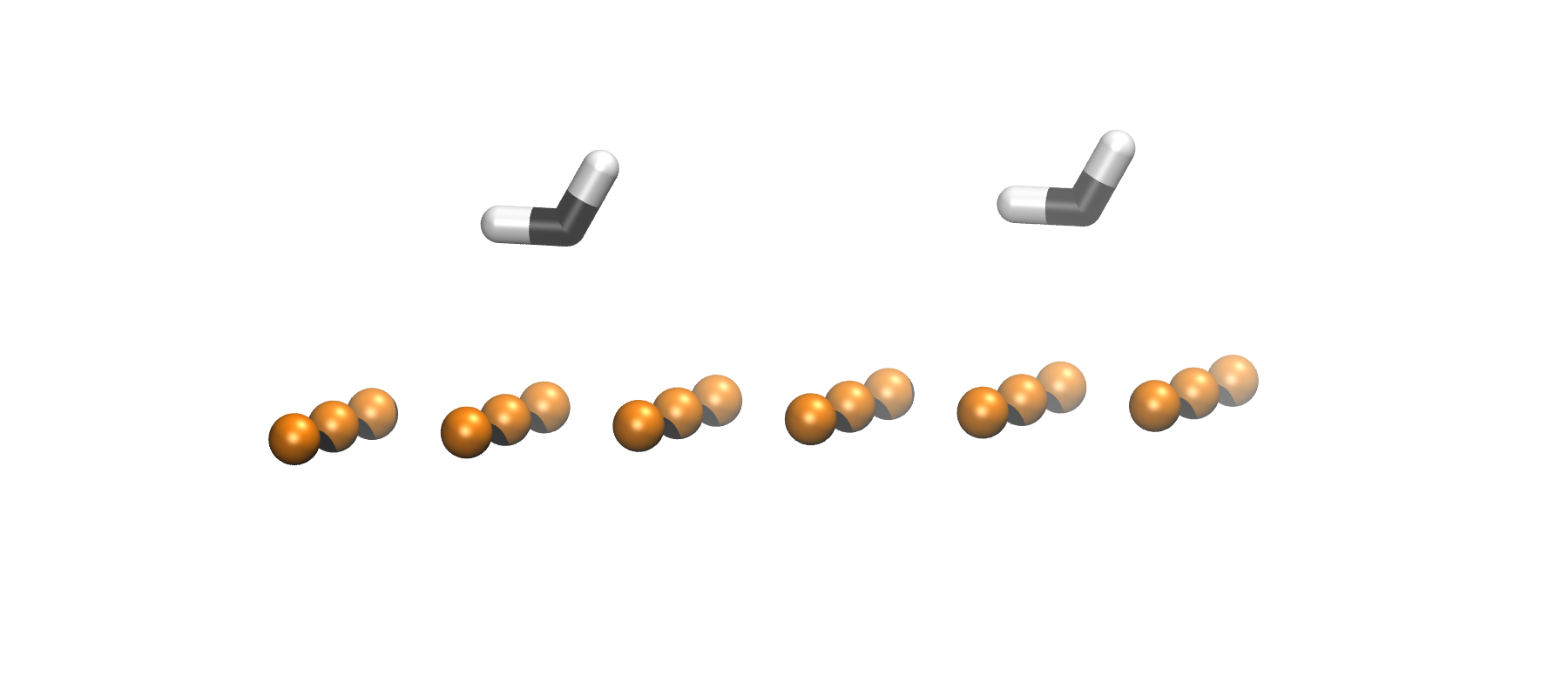}
    \includegraphics[width=0.49\linewidth]{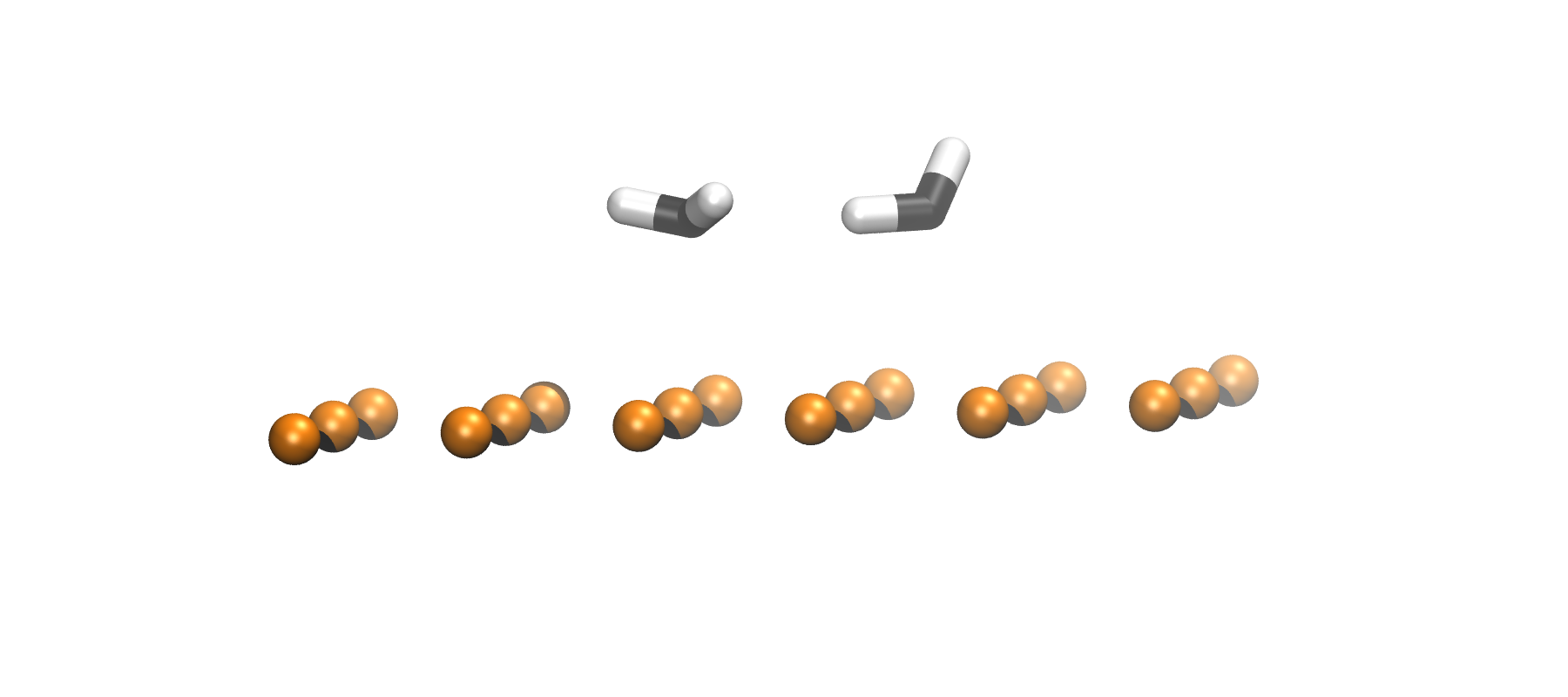}\\
    \includegraphics[width=0.49\linewidth]{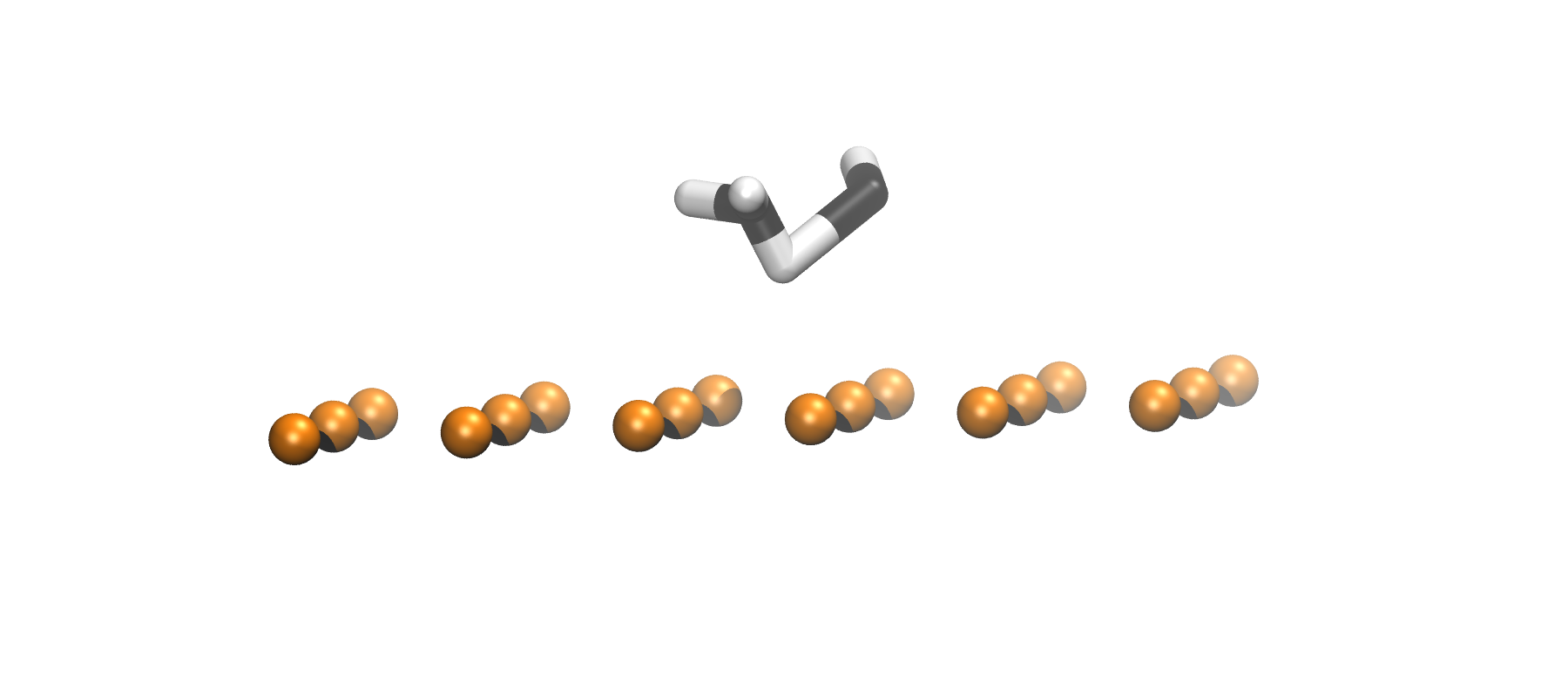}
    \includegraphics[width=0.49\linewidth]{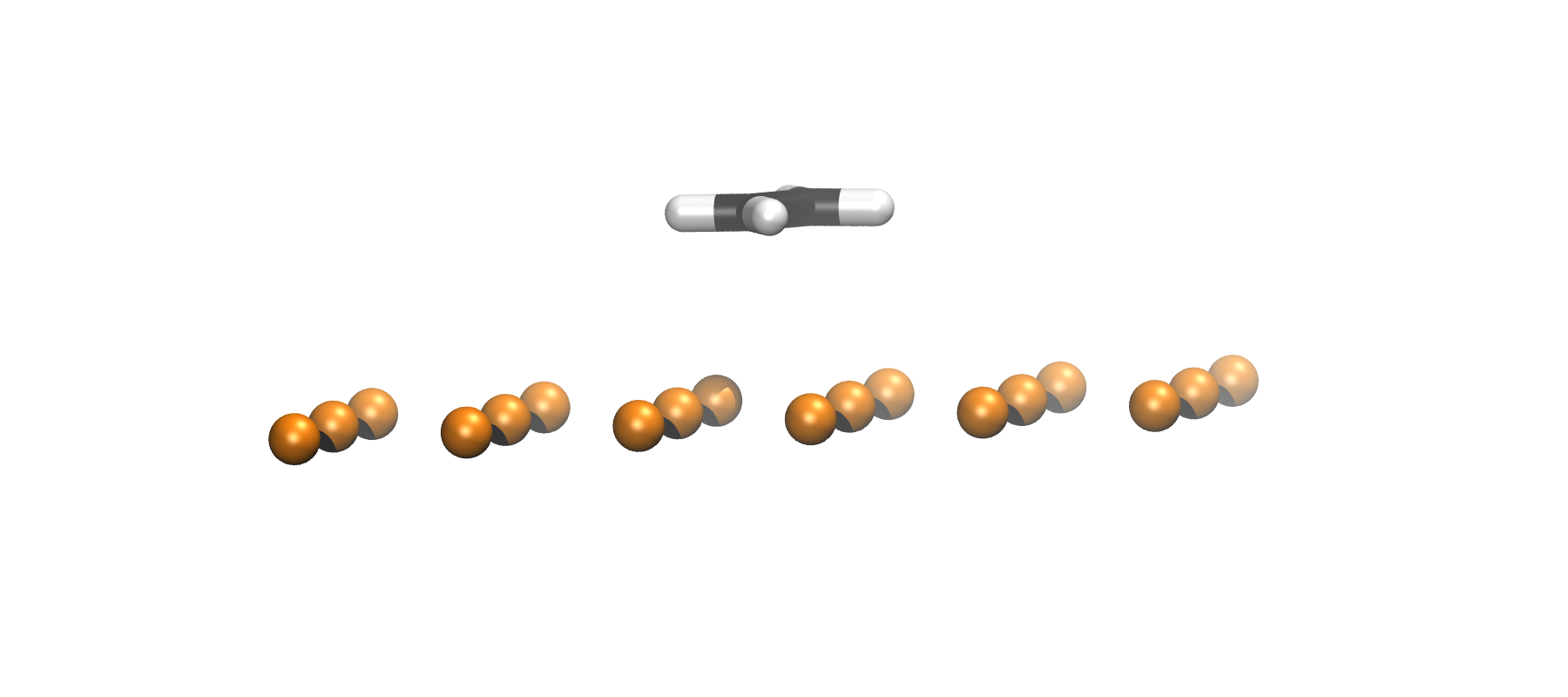}
    \caption{Snapshots at selected geometries along the reaction path on a monolayer. Top left: educts, top right: reaction coordinate index 7, bottom left: reaction coordinate index 11, bottom right: products.}
    \label{fig:path-mono}
\end{figure*}

\begin{figure*}
    \centering
    \includegraphics[width=0.49\linewidth]{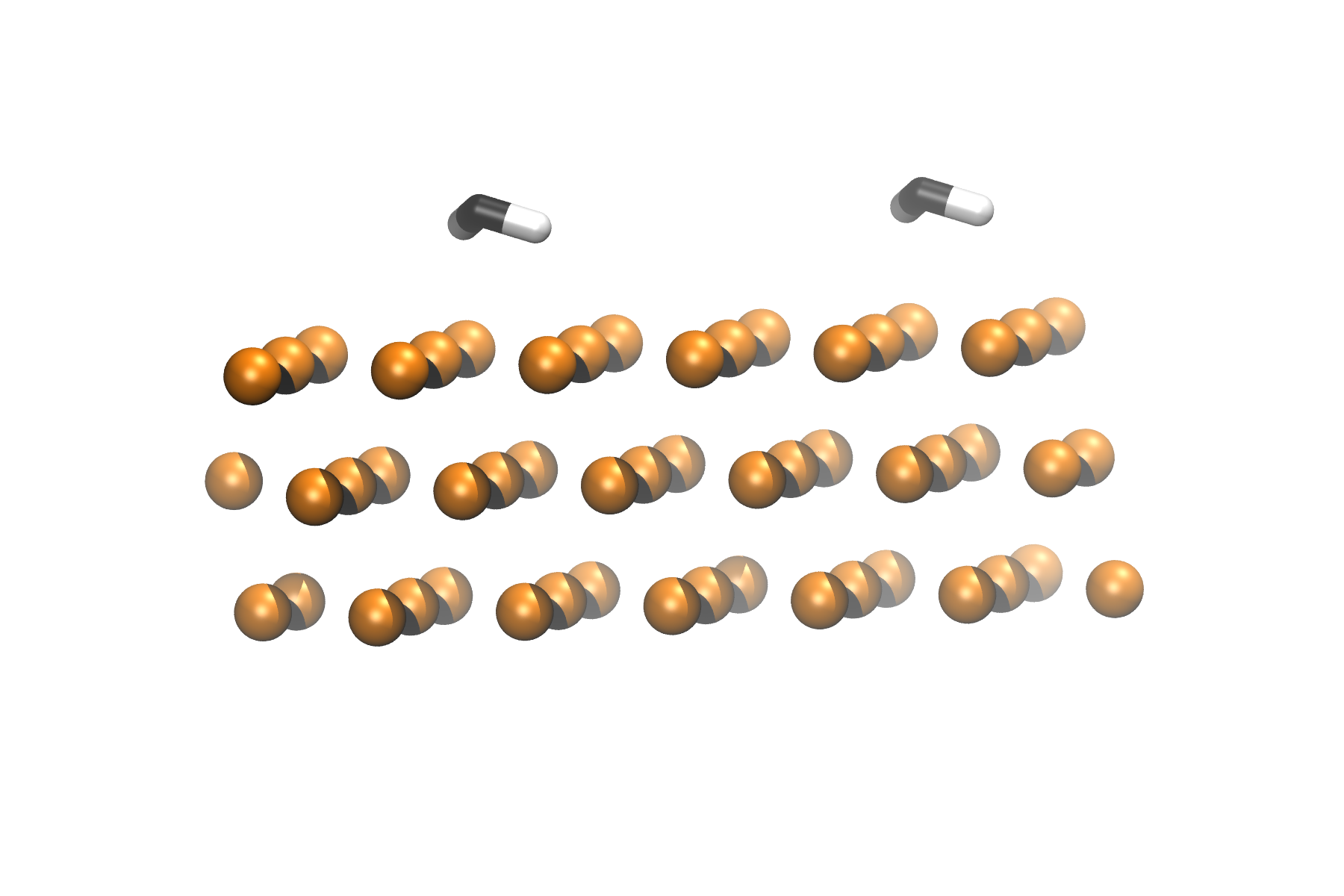}
    \includegraphics[width=0.49\linewidth]{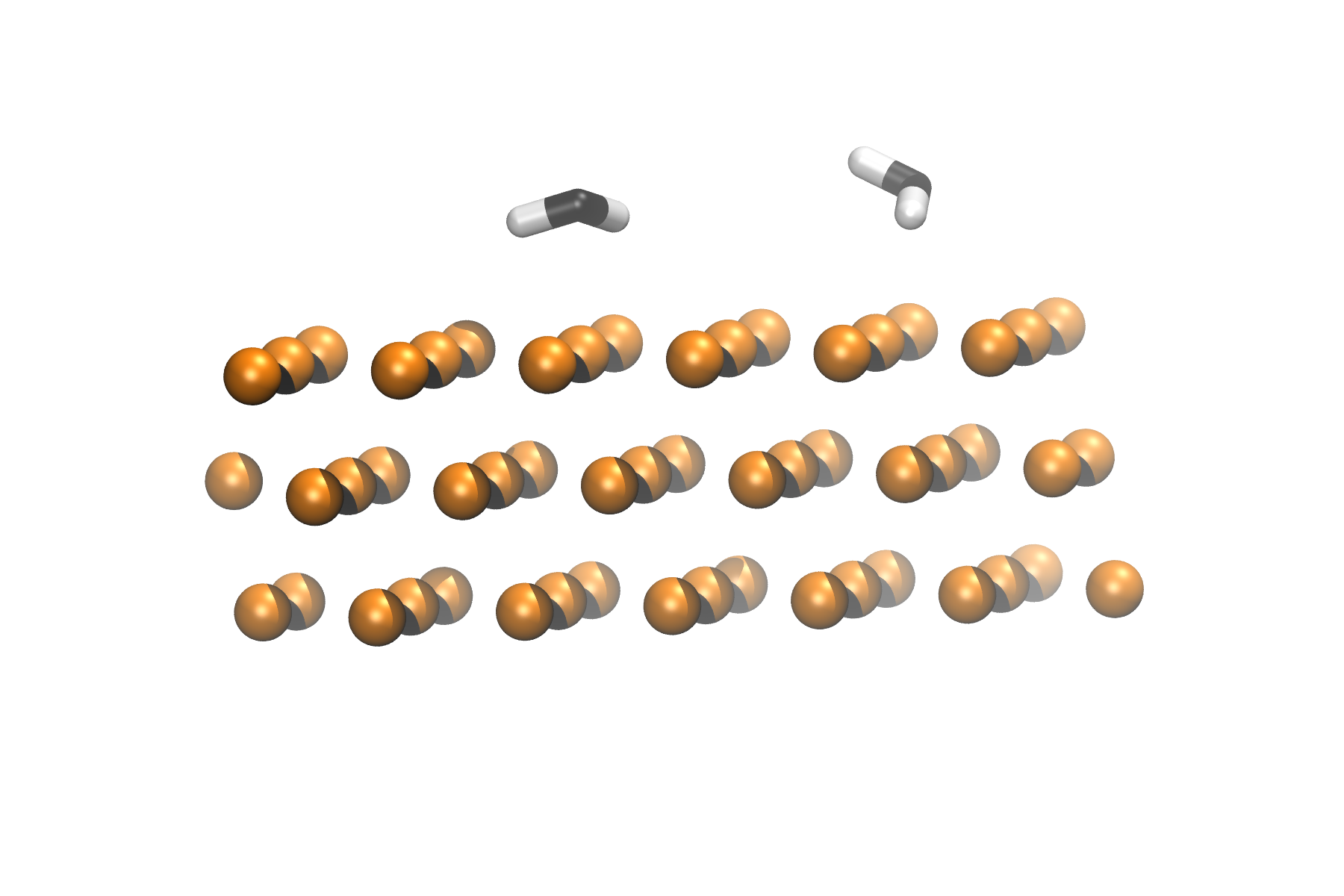}\\
    \includegraphics[width=0.49\linewidth]{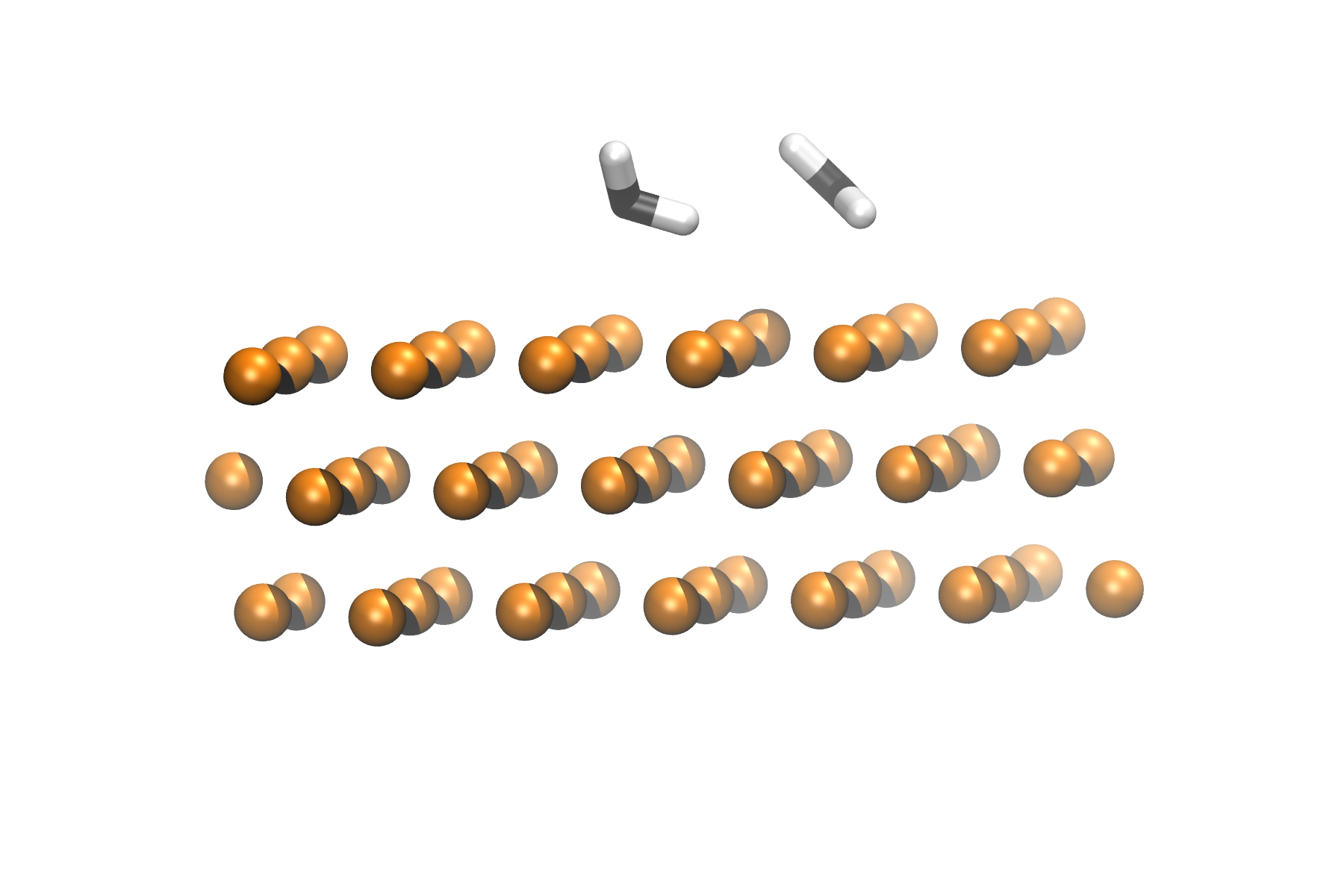}
    \includegraphics[width=0.49\linewidth]{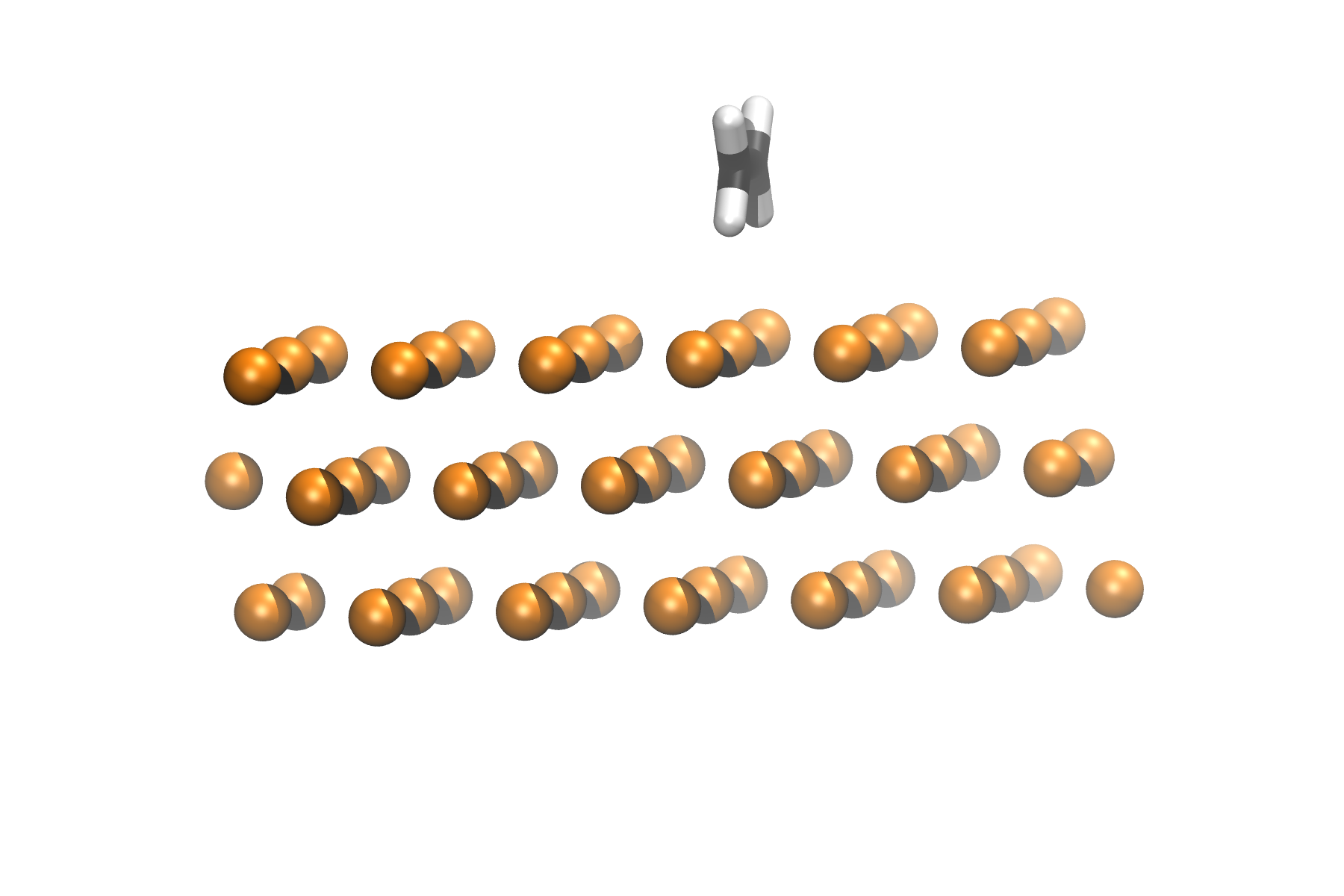}
    \caption{Snapshots at selected geometries along the reaction path on three layers of gold. Top left: educts, top right: reaction coordinate index 7, bottom left: reaction coordinate index 11, bottom right: products.}
    \label{fig:path-three}
\end{figure*}

\begin{figure}[t]
\begin{center}
\includegraphics[width=8.6cm]{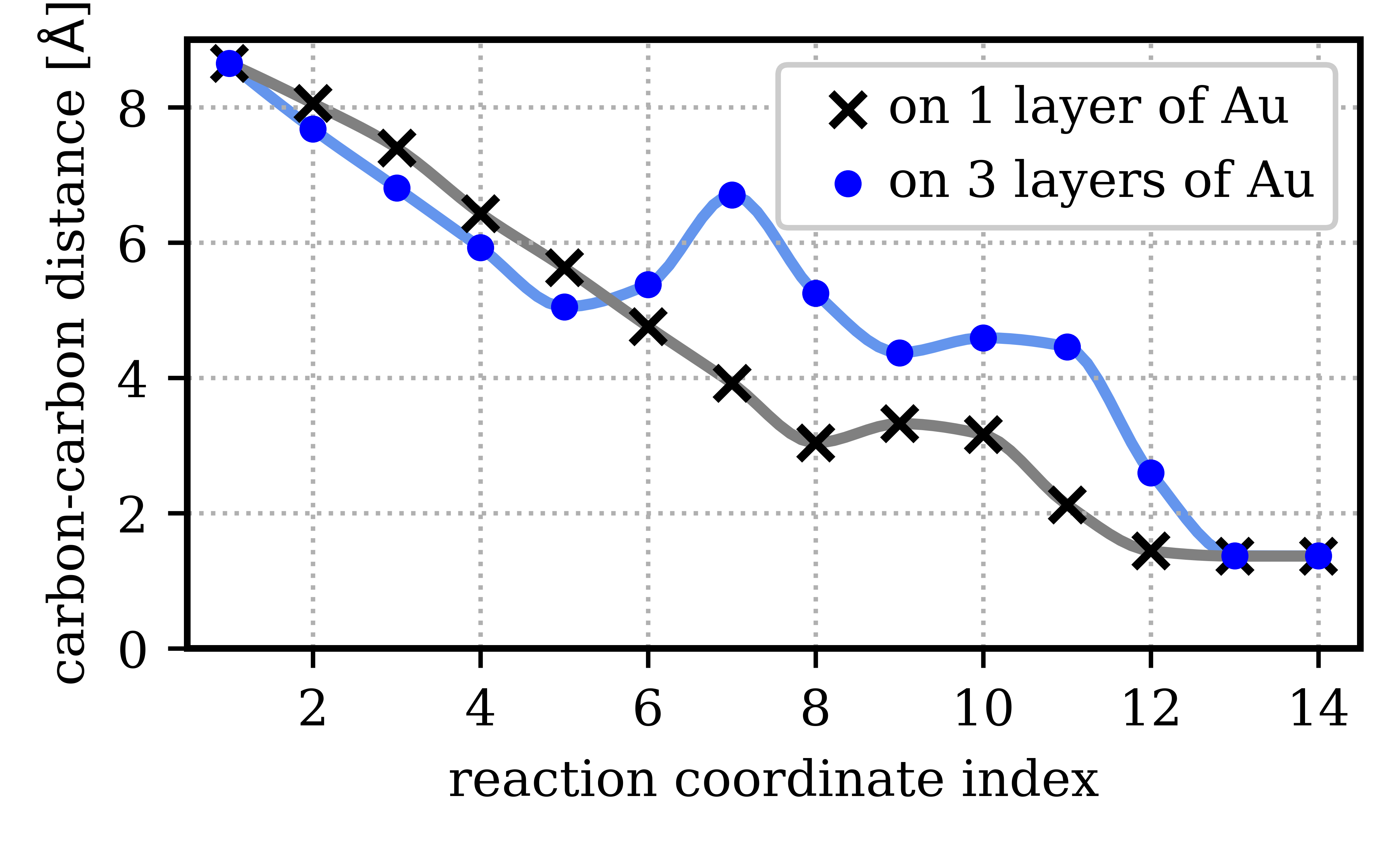}
\caption{Carbon-carbon distance of reaction paths.}
\label{ccdistance3layers}
\end{center}
\end{figure}

Qualitatively, the reaction paths on both surfaces show different behaviors.
Figures~\ref{fig:path-mono} and \ref{fig:path-three} show selected snapshots along the reaction coordinate for either system and Figure~\ref{ccdistance3layers} displays the distance between both carbon atoms along the reaction path.

The difference in adsorption energies for educts and products can be related to different configurations, as shown in Figs.~\ref{fig:path-mono} and \ref{fig:path-three}, top left and bottom right panels.
While for the gold monolayer, the methylene fragments are oriented with the carbon atom being closest to the metallic surface, the opposite is the case for three layers of gold.
The structures of the product species also show differences.
While the ethylene molecule is flat on the gold monolayer surface and an interaction between the $\pi$-electrons and the surface electrons is facilitated, on the three-layered surface the molecule is oriented perpendicular.
The different orientations on the two surfaces result in the significantly affected adsorption energies.
Apparently, surface electrons are more available for bonding and stabilization on the monolayer due to a lack of supporting gold atoms.
In contrast, on the three layers the electron density in the metallic system is higher leading to a change in the adsorbate's orientation and less stable adsorption energies for both reactants and products.

This also affects the orientation and energy profile along the reaction coordinate.
Figure~\ref{ccdistance3layers} shows the distance of both carbon atoms along the reaction path for the single gold layer (black curve) and three gold layers (blue curve).
To support the discussion of the orientation, snapshots at various reaction coordinate indices are shown in Figures S.1 and S.2 of the Supporting Information.

On the monolayer, the first eight data points along the reaction coordinate represent the diffusion of the reactant species towards each other (see Figure~\ref{fig:path-mono} top left and top right panel), which is associated with no significant change in the relative energy in Figure~\ref{energyOverReactionOn1Layers}.
Along this translational motion, only a slight rotation of the fragments is observed which leads to one of the protons being located between both carbon atoms at reaction coordinate index nine.
This is accompanied by a slight decrease in energy from reaction coordinate index eight to nine.
Subsequently, the shared proton moves towards the gold layer and the formation of the \ce{C=C} double bond is facilitated, which is seen in the further decrease of the \ce{C-C} distance and a very steep stabilization of the relative energy from reaction coordinate index nine to eleven (see Fig.~\ref{fig:path-mono} bottom left panel).
Once the product molecule is formed, a relaxation to the final minimum structure and re-orientation is observed.
This reaction on the monolayer proceeds with no significant activation energy (\SI{4.78}{\kilo\joule\per\mole}).

The reaction on three layers of gold proceeds on a different path.
At first, again the diffusion of both methylene units towards each other is observed by a decrease in the distance between both carbon atoms as shown in Fig.~\ref{ccdistance3layers}.
A comparison of Figures~\ref{fig:path-mono} and \ref{fig:path-three} shows that on three layers of gold the rotational motion of both reactants is more pronounced as compared to the monolayer, likely because of the slightly lower adsorption energy.
These rotations explain why the relative energy shown in Figure.~\ref{energyOverReactionOn3Layers} shows subtle fluctuations.

While the C-C distance shows a similar decrease until reaction coordinate five, the distance increases for the three-layer system after that and a local maximum is reached at reaction coordinate seven.
The snapshots in Figure~\ref{fig:path-three} reveals that in this part of the reaction, the translation on the surface has stopped and the fragment rotation is dominating, thus increasing the distance again while the relative energy in Fig.~\ref{energyOverReactionOn3Layers} is almost unaffected.

After that, the molecules arrange in a configuration similar to the configuration on the monolayer with one hydrogen atom being located between the carbon atoms.
However, molecules do not get to such close distances and instead, the fragments rotate further until the \ce{C=C} double bond can be formed directly.
The bond formation can be observed by a very steep decrease in the \ce{C-C} distance as well as in the energy stabilization.

\section*{Conclusion}
\label{conclusion}
	
The reaction \ce{2CH2 -> C2H4} on a gold monolayer and on three layers of gold was studied using DFT taking into account the 2D-periodic nature of this system.
We observe a strong influence of the layer thickness on the energy of adsorption as well as on the path of the reaction and the energetics.
Overall, the monolayer seems to interact more with the adsorbed species, which is probably due to the gold valence electrons being more available on the monolayer surface as compared to them being bound in the conduction band of the three-layered system.

Consequently, both educts and products are bound more stable on the thin surface.

The reaction proceeds more straightforward on the monolayer.
Translational motion of both fragments is followed by association over a shared proton until finally this proton is displaced towards the metallic layer to finally facilitate bond formation.

On three layers, in contrast, the initial translational motion is accompanied by rotational contributions.
The intermediate association is less stable and happens at larger separation of the fragments.
The rotation of the fragments dominates the majority of the reaction path until finally the double bond is formed with a very steep decrease of the \ce{C-C} distance and the relative energy.

While the findings of our study focus on rather artificial model systems, they may serve as representatives for real surface fragments, for example in dendrite structures.
The effect of the layer properties of adsorption energy, reaction path and reaction energies showcases the role of surface morphology for experiments targeting carbon upconversion.

Further follow-up studies are indicated, investigating for example the role of surfaces featuring different Miller indices, surface defects and more than three supporting gold layers.
To model dendrite structures more closely, other surface characteristics are expected to affect adsorption and reaction energetics, for example the surface curvature, which should be studied in further investigations.

\section*{Acknowledgements}

We would like to thank Fiona Jean Beck of the Australian National University for fruitful discussions.
E. P. acknowledges funding by the Deutsche Forschungsgemeinschaft (DFG, German Research Foundation) – Project-ID 398816777 – SFB 1375, Project A2.
Funding of A. S. by the Deutsche Forschungsgemeinschaft (DFG, German Research Foundation) – Project-ID 437527638 – within the GRK 2675, Project C4 is gratefully acknowledged.

\section*{Conflict of Interest}

There are no conflicts of interest.

\bibliography{EthyleneOnAu}

\includepdf[pages=-]{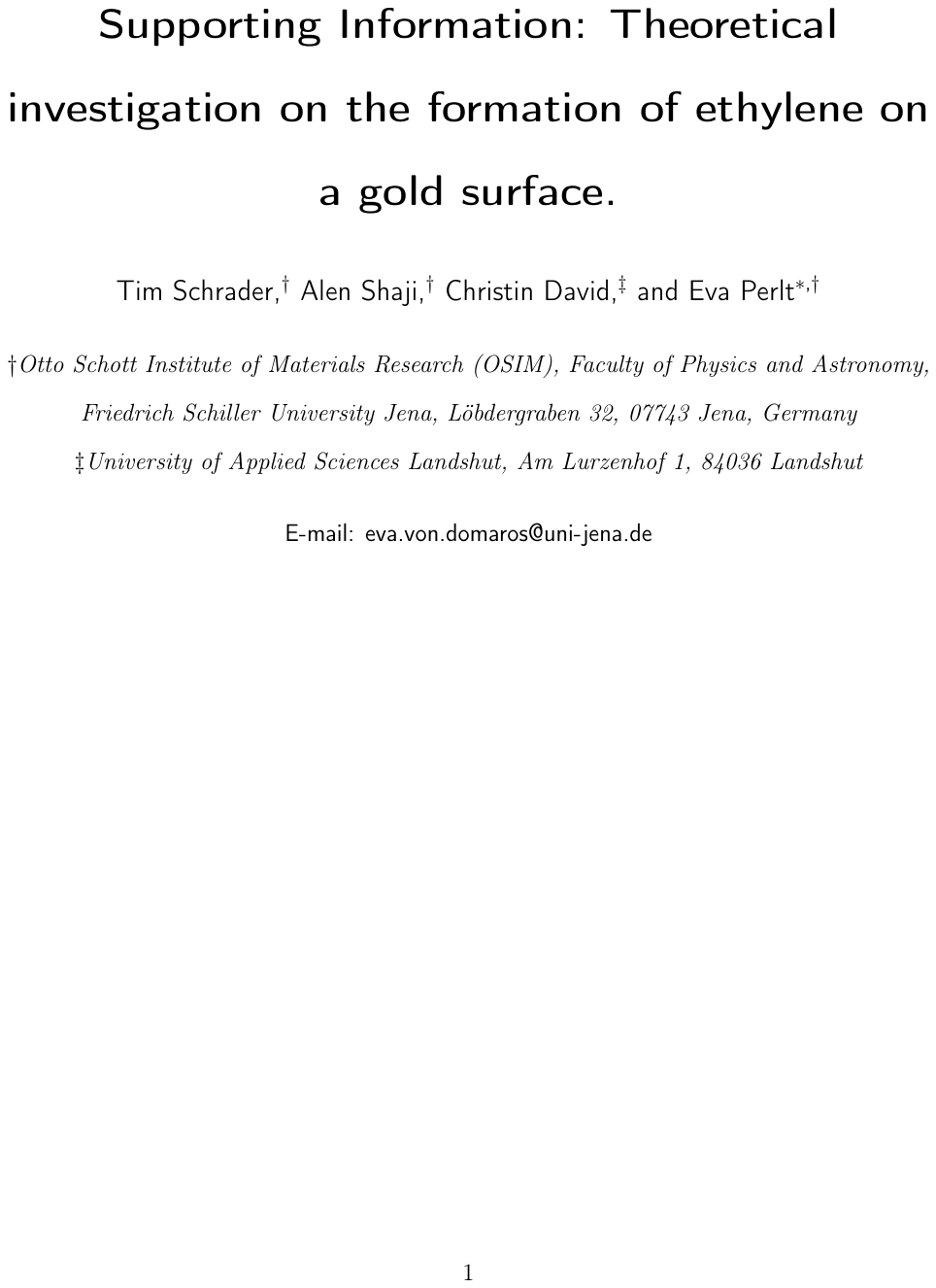}

\end{document}